\documentclass[aps,prl,twocolumn,balance,superscriptaddress,floats]{revtex4}

\pdfoutput=1
\usepackage[a4paper, total={7.2in, 10in}]{geometry}
\usepackage{latexsym}
\usepackage{dcolumn}
\usepackage{amssymb,amsmath}
\usepackage{epsf}
\usepackage{float}
\usepackage{hyperref}
\usepackage{mathrsfs}

\usepackage[pdftex]{graphicx}
\usepackage{epstopdf}
\usepackage{pdfpages}

\usepackage{upgreek}

\usepackage{xcolor}
\usepackage{soul}

\DeclareMathOperator{\sinc}{sinc}

\begin{document}


\title{Optomechanical cooling in a continuous system}

\author{Nils T. Otterstrom}
\affiliation{Department of Applied Physics, Yale University, New Haven, CT 06520 USA.}
\author{Ryan O. Behunin}
\affiliation{Department of Applied Physics, Yale University, New Haven, CT 06520 USA.}
\affiliation{Department of Physics and Astronomy, Northern Arizona University, Flagstaff, AZ 86001 USA.}
\author{Eric A. Kittlaus}
\affiliation{Department of Applied Physics, Yale University, New Haven, CT 06520 USA.}
\author{Peter T. Rakich}
\affiliation{Department of Applied Physics, Yale University, New Haven, CT 06520 USA.}


\date{\today}

\begin{abstract}

Radiation-pressure-induced optomechanical coupling permits exquisite control of micro- and mesoscopic mechanical oscillators.  This ability to manipulate and even damp mechanical motion with light---a process known as dynamical backaction cooling---has become the basis for a range of novel phenomena within the burgeoning field of cavity optomechanics, spanning from dissipation engineering to quantum state preparation. As this field moves toward more complex systems and dynamics, there has been growing interest in the prospect of cooling traveling-wave phonons in continuous optomechanical waveguides.  Here, we demonstrate optomechanical cooling in a continuous system for the first time.  By leveraging the dispersive symmetry breaking produced by inter-modal Brillouin scattering, we achieve continuous mode optomechanical cooling in an extended 2.3-cm silicon waveguide, reducing the temperature of a band of traveling-wave phonons by more than 30 K from room temperature. This work reveals that optomechanical cooling is possible in macroscopic linear waveguide systems without an optical cavity or discrete acoustic modes.  Moreover, through an intriguing type of wavevector-resolved phonon spectroscopy, we show that this system permits optomechanical control over continuously accessible groups of phonons and produces a new form of nonreciprocal reservoir engineering. Beyond this study, this work represents a first step towards a range of novel classical and quantum traveling-wave operations in continuous optomechanical systems.

\end{abstract}

\maketitle

\section{Introduction}

The ability to control and harness optical forces within mesoscale systems has enabled a range of cavity-optomechanical devices as the basis for numerous classical and quantum operations  \cite{aspelmeyer2014cavity}. Integral to these developments is a technique called sideband cooling, in which dynamical backaction is used to produce a net cooling effect on a mechanical oscillator \cite{braginskii1970investigation,cohadon1999cooling,teufel2011sideband,chan2011laser}. In the framework of cavity optomechanics, this is accomplished using an optical cavity to enhance and manipulate optomechanical coupling to discrete phonon modes. This strategy for optomechanical cooling is central to a host of novel functionalities and dynamics, ranging from precision metrology \cite{ap:schliesser2009resolved} to quantum state generation \cite{ap:o2010quantum,ap:verhagen2012quantum,ap:palomaki2013entangling} and fundamental tests of quantum decoherence \cite{ap:marshall2003towards}.  Beyond single-mode cavity optomechanics, intriguing opportunities are presented by extended optomechanical systems that possess many degrees of freedom. For example, multimode optomechanical devices \cite{jayich2008dispersive,bhattacharya2008multiple, schmidt2012optomechanical,lee2015multimode}, optomechanical arrays \cite{heinrich2011collective,xuereb2012strong,zhang2012synchronization,tomadin2012reservoir,schmidt2015optomechanical}, and waveguide-coupled resonators \cite{safavi2011proposal,habraken2012continuous,fang2016optical,patel2017single} offer new strategies for everything from reservoir engineering \cite{tomadin2012reservoir,metelmann2015nonreciprocal,seif2018thermal} to quantum networking on a chip \cite{schmidt2012optomechanical,safavi2011proposal,habraken2012continuous,fang2016optical,patel2017single}.

An intriguing limiting case of these systems is known as continuum optomechanics \citep{rakich2016quantum}, in which optical fields are used to control and manipulate sound waves in a translationally invariant medium \cite{rakich2016quantum,van2016unifying}.  These extended, optically transparent systems give rise to a continuum of optically addressable acoustic states \cite{stiller2018crosstalk}, and could open the door to novel forms of squeezing \cite{schmidt2012optomechanical}, optomechanical quantum networks \cite{habraken2012continuous,shumeiko2016quantum,vermersch2017quantum}, and quantum nonlinear optics \cite{zoubi2017quantum} with unprecedented optical and acoustic bandwidth.  An important step towards these novel operations is the ability to selectively heat or cool traveling-wave phonon fields within continuous optomechanical waveguides.

As a form of distributed optomechanical coupling, Brillouin interactions could provide a promising avenue for controlling groups of phonons in continuous systems, and have recently been proposed as a potential method for achieving continuum optomechanical cooling \cite{chen2016brillouin,huy2017brillouin}.  While spontaneous Brillouin scattering has been used to cool phonon modes within discrete cavity-optomechanical systems \cite{bahl2012observation}, Brillouin cooling in a continuous system has yet to be demonstrated.  Recent theoretical work has proposed the use of high-gain Brillouin-active waveguides to achieve continuum optomechanical cooling \cite{chen2016brillouin}. However, this analysis suggested that cooling of this type might be beyond the reach of most experimental systems.

\begin{figure*}[hbtp]
\centering
\includegraphics[width=\linewidth]{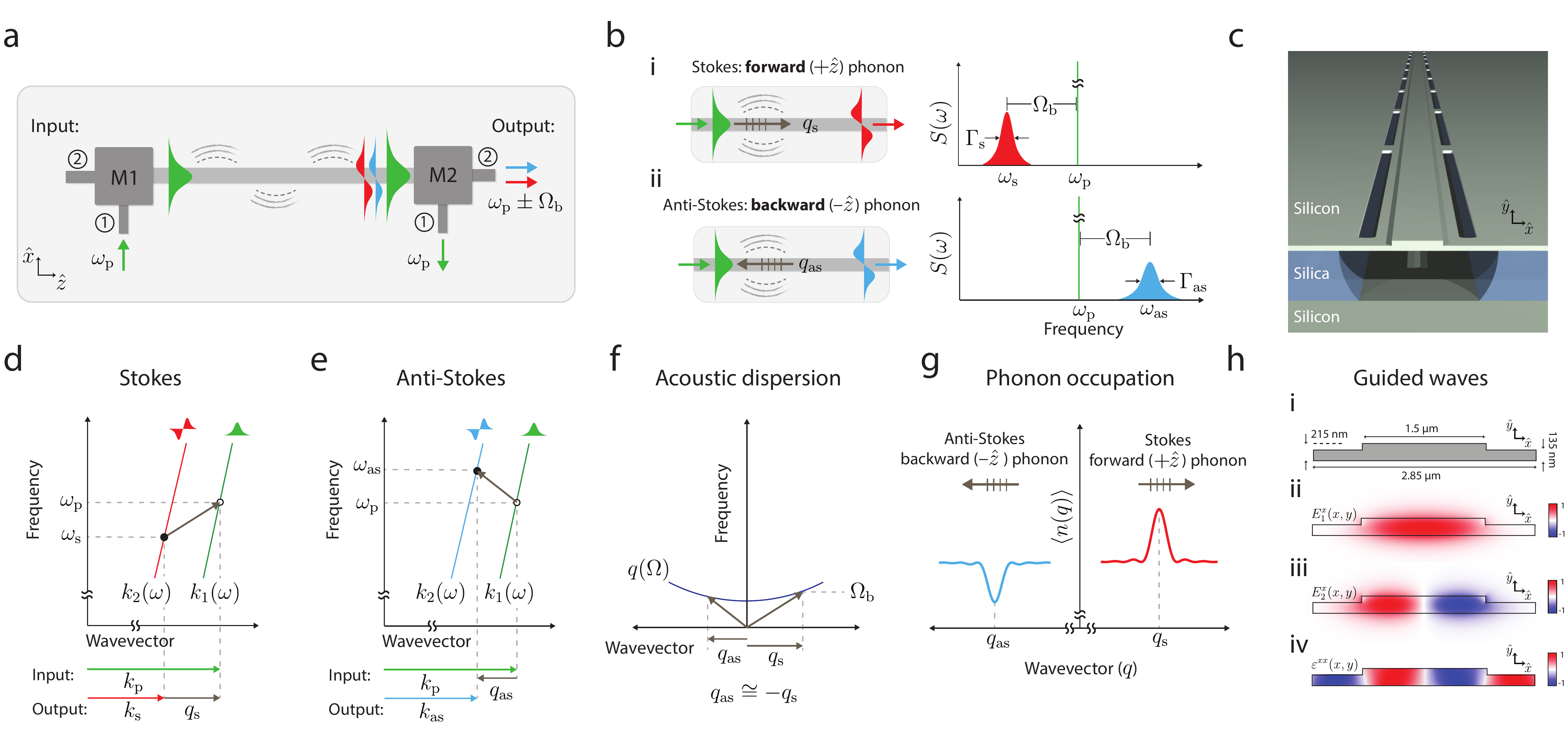}
\caption{Panel (a) illustrates the waveguide system and basic operation scheme.  Probe light (green) of frequency $\omega_{\rm p}$ is coupled into the symmetric spatial mode of a Brillouin-active silicon waveguide through an integrated mode multiplexer (M1). This light interacts with forward- and backward-propagating thermal phonon fields, which produce respective Stokes (red) and anti-Stokes (blue) sidebands that propagate in the antisymmetric spatial mode.  An integrated mode multiplexer (M2) then demultiplexes the scattered light for spectral analysis. (b) illustrates the optical Stokes and anti-Stokes spectra produced by spontaneous inter-modal Brillouin scattering due to (bi) forward- and (bii) backward-propagating phonons, respectively.  The spectral width of each sideband reveals the temporal dissipation rates and lifetimes of the phonons participating in  the Stokes and anti-Stokes processes. Panel (c) depicts the suspended silicon ridge waveguide that guides both optical and acoustic waves. Panels (d)-(f) illustrate the energy conservation and phase-matching requirements for these spontaneous Brillouin processes. (d)-(e) plot the optical dispersion relations for the symmetric and antisymmetric optical modes as well as the phonons that mediate the Stokes and anti-Stokes scattering. In the Stokes process, the phonon that mediates scattering from initial state (open circle) to the final state (closed circle) is a forward-propagating field.  By contrast, as diagrammed in (e), phase-matching dictates that the phonon responsible for anti-Stokes scattering must be a backward-propagating wave.  These two phonons must satisfy the acoustic dispersion relation for the Lamb-like acoustic mode that mediates inter-modal scattering, as shown in (f). In this system, the Stokes and anti-Stokes phonons are essentially degenerate in both Brillouin frequency and wavevector magnitude but propagate in opposite directions. (g) This spontaneous process simultaneously reduces the thermal occupation $\left<n(q)\right>$ of phase-matched backward-propagating phonons and increases that of the phase-matched forward propagating phonons. As a result, an incident laser field drives the average momentum of the thermal bath of phonons out of equilibrium, producing a net phonon flux. (hi) diagrams the cross-sectional geometry of the hybrid photonic-phononic waveguide. Panels (hii) and (hiii) plot the $x$-polarized component of the TE-like symmetric ($E^{x}_1(x,y)$) and antisymmetric ($E^{x}_2(x,y)$) simulated mode profiles, respectively. (hiv) shows the simulated strain profile of the 6-GHz elastic mode that mediates spontaneous inter-modal scattering. Here, we have plotted $\varepsilon^{xx}$, which is the dominant component in the inter-modal acousto-optic coupling.}
\label{fig:deviceconcept}
\end{figure*}

In this Letter, we demonstrate phonon cooling in a continuous optomechanical system for the first time. Using a phase-matched Brillouin process in a multimode optomechanical waveguide, we are able to selectively address and cool phonons within a continuous band of accessible states---without the need for an optical or acoustic resonator. Due to the wavevector-selective nature of this process, 
the inter-band Brillouin coupling in this macroscopic (cm-scale) waveguide system is sufficient to reduce the temperature of a band of traveling-wave phonons by more than 30 K from room temperature. Leveraging this phase-matched interaction, we perform wavevector-resolved phonon spectroscopy and demonstrate wavevector-tunable phonon control. In this way, we are able to selectively probe and cool continuously accessible groups of phonons, simply by tuning the wavelength of the incident light. In addition, we develop a succinct theoretical framework to understand our observations and present general guidelines for continuum optomechanical cooling in extended waveguide systems. Since this continuous system does not possess discrete acoustic modes, we show that this type of cooling can be viewed as a form of wavevector-selective reservoir engineering that yields nonreciprocal phonon transport.

\section{Results}

We demonstrate continuous-mode phonon cooling by leveraging a guided-wave optomechanical process, termed inter-modal Brillouin scattering \cite{kang2010all,kittlaus2017chip}, within a 2.3 cm-long photonic-phononic waveguide. This optomechanical silicon waveguide is fabricated from a single-crystal silicon-on-insulator (SOI) wafer (for more information see Supplementary Section 5). Throughout the device, light is guided by total internal reflection using a ridge waveguide structure, which supports low-loss guidance of TE-like symmetric and anti-symmetric spatial modes (see Fig. \ref{fig:deviceconcept}hii-iii), with propagation constants given by $k_1(\omega)$ and $k_2(\omega)$, respectively.  By removing the oxide undercladding, the suspended interaction regions (see Fig. \ref{fig:deviceconcept}c) also support a 6-GHz guided elastic wave, which mediates efficient ($\rm G_{\rm b}\cong470$ $ \rm W^{-1}m^{-1}$) nonlinear coupling between the two optical modes. This phonon field has an intrinsic dissipation rate ($\Gamma/(2 \pi)$) of 14.2 MHz, corresponding to a decay length ($l_{c}$) of $60$ $\upmu$m.  The cross section of this device is designed for maximal inter-modal Brillouin coupling, with transverse dimensions identical to the device studied in Ref. \cite{kittlaus2017chip}.   This hybrid photonic-phononic waveguide structure is continuously suspended by an array of nano-scale tethers, permitting seamless traveling-wave coupling over centimeter length scales.

\begin{figure*}[hbtp]
\centering
\includegraphics[width=.7\linewidth]{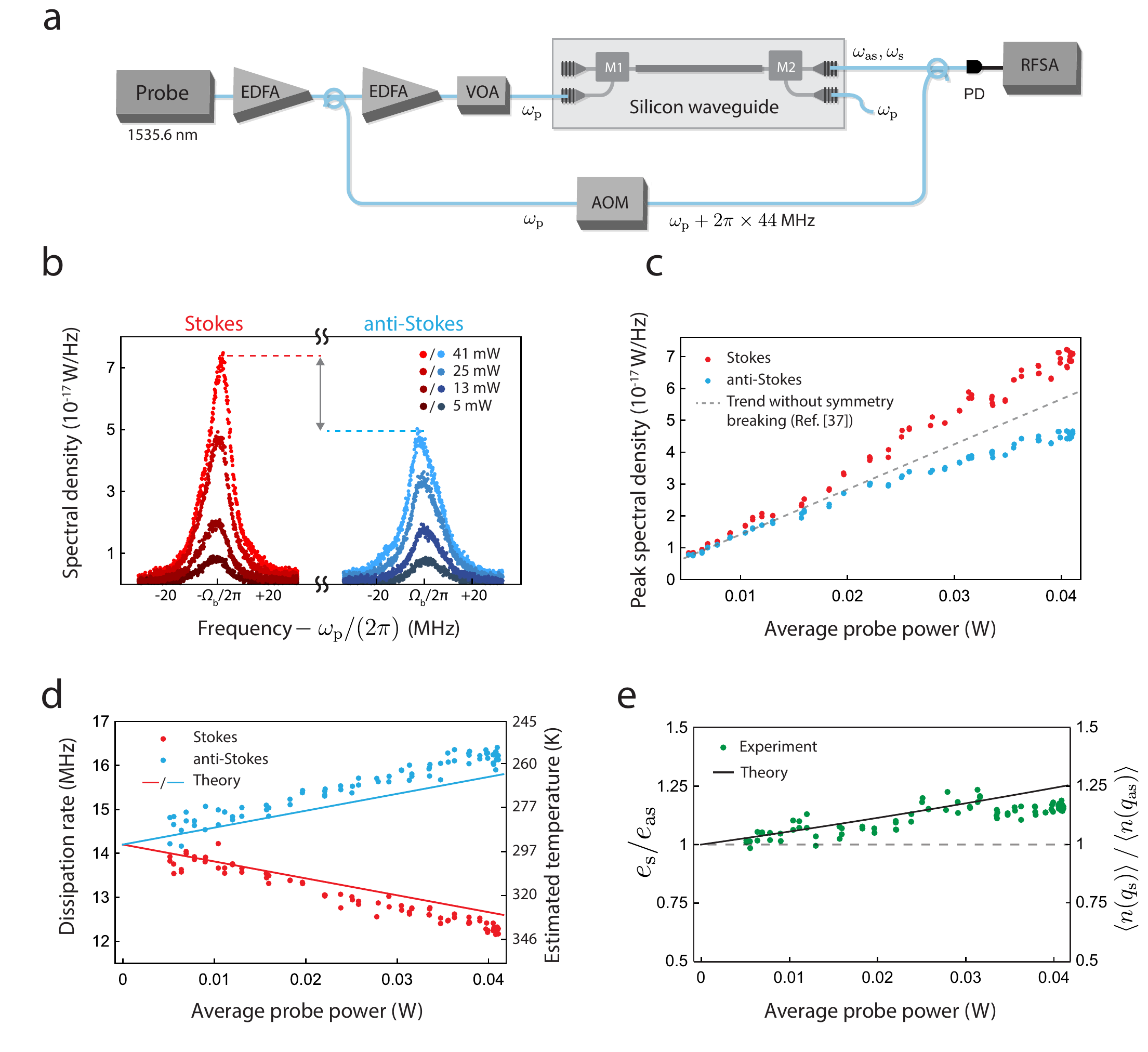}
\caption{Experimental heterodyne setup and measurements of spontaneous inter-modal Brillouin scattering. Panel (a) diagrams the heterodyne scheme used to modify and probe the phonon dynamics. Experiments are performed at room temperature and atmospheric pressure. A cw laser source (vacuum wavelength of 1535.5 nm) is used to synthesize a strong probe wave (upper arm) and a 44-MHz blue-shifted optical local oscillator for heterodyne detection (lower arm). The probe wave intensity is controlled using an erbium-doped fiber amplifier (EDFA) and a variable optical attentuator (VOA) before being coupled on-chip, where it interacts with a thermal phonon field via inter-modal Brillouin scattering. The scattered light propagates in the antisymmetric spatial mode and then is combined with the optical local oscillator (LO).  The interference of the Stokes and anti-Stokes waves with the optical LO on a photo-receiver produces unique microwave spectra centered at $\Omega_{\rm b}+2 \pi\times44$MHz and $\Omega_{\rm b}-2 \pi\times 44$MHz, respectively.  Panel (b) plots a series of Stokes and anti-Stokes heterodyne spectra at four distinct probe powers. Note the power-dependent asymmetry between the Stokes and anti-Stokes spectra in both the peak spectral density and spectral width. (c) plots the peak spectral density of these spectra as a function of probe power. The trend curve (from Ref. \cite{kharel2016noise}) does not take into account the spontaneous dispersive symmetry breaking, but does provide a first-order benchmark for the observed Stokes and anti-Stokes powers.  Panel (d) plots the fitted linewidths of the Stokes and anti-Stokes spectra. The temperature is estimated using the model derived in Supplementary Section 1.  These measurements reveal that the collective lifetime of the anti-Stokes phonons is reduced while that of the Stokes phonons is enhanced.  At a maximum probe power of 42 mW, this asymmetry in spectral width corresponds to more than 30 K of cooling/heating from room temperature. (e) shows the relative scattering efficiencies of the Stokes and anti-Stokes processes (and respective phonon occupations) as the probe power is increased.}
\label{fig:singleprobe}
\end{figure*}

We realize optomechanical cooling in this continuous waveguide system through spontaneous Brillouin scattering.  Probe light of frequency $\omega_{\rm p}$ (green in Fig. \ref{fig:deviceconcept}a) is coupled in the symmetric optical spatial mode of the multimode optomechanical waveguide, which interacts with thermally driven phonons through spontaneous inter-modal Brillouin scattering. Forward- and backward-propagating phonons produce Stokes and anti-Stokes sidebands, respectively. Both of these scattered waves propagate in the antisymmetric spatial mode of the Brillouin-active waveguide (Fig. \ref{fig:deviceconcept}a-b). 

In this system, inter-modal Brillouin scattering produces a form of phase-matching-induced symmetry breaking---or decoupling between the Stokes (heating) and anti-Stokes (cooling) processes---that is quite distinct from that of sideband cooling in cavity-optomechanical systems. This symmetry breaking arises because the Stokes and anti-Stokes processes are mediated by distinct groups of phonons that propagate in opposite directions; specifically, phase-matching requires that $ q_{\rm s}(\Omega_{\rm b}) = k_1(\omega_{\rm p}) - k_2(\omega_{\rm s})$ and $q_{\rm as}(\Omega_{\rm b}) = k_2(\omega_{\rm as})-k_1(\omega_{\rm p})$, where $\omega_{\rm s}$ and $\omega_{\rm as}$ are the respective Stokes and anti-Stokes frequencies, and $q_{\rm s}(\Omega_{\rm b})$ and $q_{\rm as}(\Omega_{\rm b})$ are the respective propagation constants of the Stokes and anti-Stokes phonons at the Brillouin frequency $\Omega_{\rm b}$. As a result, the Stokes process is mediated by a forward-propagating acoustic field while the anti-Stokes process is mediated by a backward-propagating acoustic field, as illustrated in Fig. \ref{fig:deviceconcept}d-f (for further discussion see Ref. \cite{kittlaus2017chip}). Thus, phase-matched forward-traveling phonons experience heating while phase-matched backward-traveling phonons experience cooling (see Fig. \ref{fig:deviceconcept}b,g). Note that these behaviors are in contrast to those of forward intra-modal Brillouin processes, which do not exhibit phase-matching-induced symmetry breaking or produce cooling \cite{van2017thermal}. After traversing the Brillouin-active waveguide, the probe and spontaneously-generated sidebands are demultiplexed using a integrated mode multiplexer, which routes the Stokes and anti-Stokes light off-chip for spectral analysis, as illustrated in Fig. \ref{fig:deviceconcept}a.  As shown in Fig. \ref{fig:deviceconcept}b, the spectral width of each sideband ($\Gamma_{\rm s}$, $\Gamma_{\rm as}$) gives a direct measure of the associated dissipation rates and lifetimes of the phonons that mediate the Stokes and anti-Stokes processes, respectively \cite{kharel2016noise}.

Through this phase-matched process, spontaneous inter-modal Brillouin scattering directly modifies the thermodynamic state of the traveling-wave Stokes and anti-Stokes phonon fields. For the phase-matched backward-propagating (anti-Stokes) phonons, the presence of a strong probe field produces an additional damping mechanism; phonons annihilated by this scattering process are converted to anti-Stokes photons, which escape the system at a rate much greater than the intrinsic dissipation rate of the phonon field. As a result, the anti-Stokes process reduces the average lifetime and occupation of the phonon field; cooling of the anti-Stokes phonons is manifest as both a reduction in the phonon lifetime---or broadening of the spontaneous linewidth ($\Gamma_{\rm as}$)---and a decrease of the anti-Stokes scattering efficiency ($e_{\rm as}$). By contrast, the presence of a strong probe field yields linewidth narrowing of the Stokes sideband and increases the Stokes scattering efficiency ($e_{\rm s}$).  These two, intrinsically-decoupled processes occur simultaneously, heating forward-propagating phonons and cooling backward-propagating phonons in this continuous waveguide system (see Fig. \ref{fig:deviceconcept}g). Under these conditions, the forward- and backward- propagating phonon fluxes are no longer balanced, yielding an unusual form of nonreciprocal phonon  transport (see Fig. \ref{fig:deviceconcept}g).

\subsection{Observation of Brillouin cooling}

We examine the effects of Brillouin cooling by performing optical heterodyne spectroscopy on the spontaneously scattered light.  Figure \ref{fig:singleprobe}a diagrams the experimental setup used for these measurements, which are conducted at room temperature and atmospheric pressure.  Probe light of wavelength 1535.5 nm is generated by a continuous wave (cw) tunable external cavity laser and split along two paths.  One arm synthesizes an optical local oscillator (LO) for heterodyne detection by passing the light through an acousto-optic frequency shifter, which blue-shifts the light by $\Delta = 2 \pi \times 44 $ MHz.  In the second arm, the probe wave intensity is controlled using an erbium-doped fiber amplifier (EDFA) and a variable optical attenuator (VOA) before being coupled on chip. Following the Brillouin-active waveguide, the spontaneously scattered light is separated from the probe wave using an integrated mode multiplexer and routed off-chip, where it is combined with the optical local oscillator for spectral analysis.  By allowing the Stokes and anti-Stokes light to interfere with the LO on a fast photo-receiver, the optical signals of interest are converted to distinct microwave tones (of frequency $\Omega_{\rm b}+\Delta$ and $\Omega_{\rm b}-\Delta$, corresponding to Stokes and anti-Stokes optical frequencies of $\omega_{\rm p}-\Omega_{\rm b}$ and $\omega_{\rm p}+\Omega_{\rm b}$, respectively), which can then be rapidly recorded using a radio-frequency spectrum analyzer.  In this way, heterodyne spectroscopy permits sensitive, frequency-resolved, high-resolution spectral measurements of Stokes and anti-Stokes sidebands simultaneously.  Since the scattered Stokes and anti-Stokes waves are produced by a linear scattering between probe and phonon fields through a spontaneous process (see Supplementary Section 1), this heterodyne measurement reveals the spectra of the phase-matched phonons.

From these measurements, we observe a power-dependent asymmetry in both scattering efficiency and linewidth of the spontaneously-generated Stokes and anti-Stokes sidebands, in agreement with our theoretical predictions (see Supplementary Section 1).  Figure \ref{fig:singleprobe}b plots a series of heterodyne Stokes and anti-Stokes spectra at various probe powers.  At low powers, the Stokes and anti-Stokes spectra exhibit nearly identical scattering rates and linewidths.  However, as the probe power increases, the peak spectral density of the Stokes wave increases at a rate greater than that of the anti-Stokes. In addition, the anti-Stokes (Stokes) spectrum exhibits appreciable spectral broadening (narrowing), revealing a Brillouin-induced reduction (increase) in phonon lifetime (see Fig. \ref{fig:singleprobe}d).  This linewidth broadening corresponds to a reduction in anti-Stokes phonon population by approximately 11\%, or 32 K of effective cooling relative to room temperature.  
\begin{figure*}[hbtp]
\centering
\includegraphics[width=.85\linewidth]{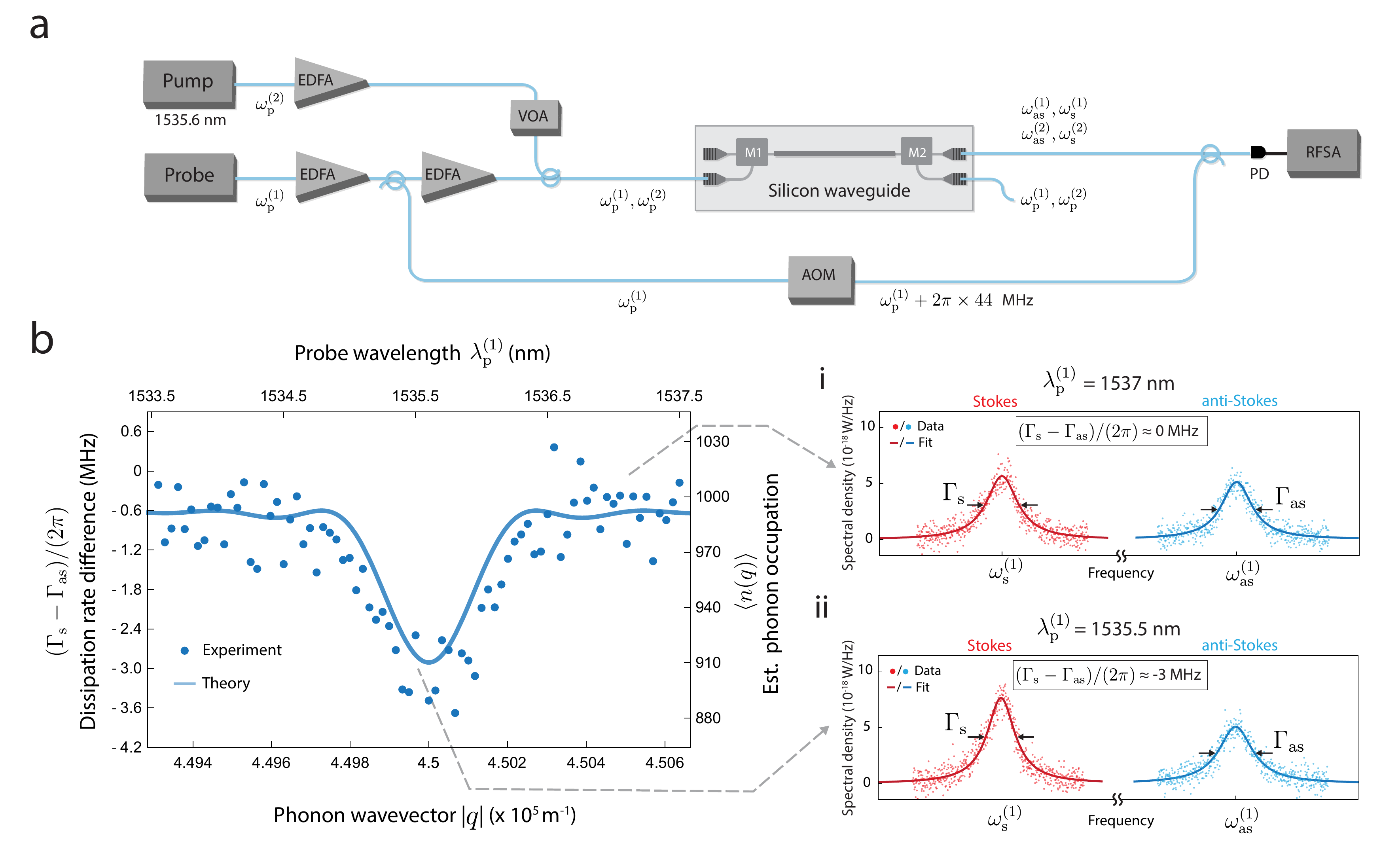}
\caption{Pump-probe spontaneous Brillouin scattering experimental setup and measurements. Panel (a) diagrams the experimental scheme used for the pump-probe experiments.  These measurements involve two tunable, cw lasers. The first, labeled the probe laser (indexed by superscript $^{(1)}$), which has a fixed power of 8 mW, is used to probe the phonon dynamics.  Since the optical LO (blue-shifted 44 MHz by the acousto-optic modulator (AOM)) is synthesized from the probe source, the heterodyne Stokes and anti-Stokes signals centered at ($\Omega_{\rm b}+2 \pi\times44$ MHz and $\Omega_{\rm b}-2 \pi\times 44$ MHz, respectively) originate entirely from the spontaneously scattered probe light.  The second source, labeled the pump laser (indexed by superscript $^{(2)}$), is used to investigate the effect of a strong pump wave on the phonon dynamics.
Panel (b) shows the difference in Stokes and anti-Stokes phonon dissipation rates (spectral widths of the Stokes and anti-Stokes sidebands produced by the probe laser) as a function of probe wavelength while the pump wavelength remains fixed at $\lambda^{(2)}_{\rm p} = 1535.6$ nm. Each data point represents the difference between Stokes and anti-Stokes dissipation rates at a pump power ($P_{\rm p}^{(2)}$) of 30 mW, obtained by fitting the measured spectral widths over a series of 10 different pump powers.
The theoretical trend (see Eq. \ref{eq:occupff}) is superimposed. The phonon wavevector is calculated from the effective phase and group indices of the two optical spatial modes supported by the silicon waveguide (see Supplementary Section 3,6), and the anti-Stokes phonon occupation is estimated by a comparison with the spatio-temporal theory. This comparison is verified by additional high-resolution pump-probe experiments (see Supplentary Section 4). (bi) Example spectra when the probe wave is not phase-matched to the same group of phonons as the pump wave. In this case, the dissipation rates for the Stokes and anti-Stokes phonons remain constant as the pump power increases. (bii) Example spectra when the probe wave is phase-matched to the same group of phonons as the pump wave. Here, increasing the pump power enhances the dissipation rate asymmetry (see also Supplementary Section 4).  These data reveal that the pump wave reduces the phonon occupation over a narrow band of phonon wavevectors, with a bandwidth given by $\Delta q =2.78/L$ (see Supplementary Section 3).  }
\label{fig:pumpprobe}
\end{figure*}

\subsection{Pump-probe experiments}

An intriguing aspect of this system is the ability to optically address and control groups of phonons within a continuous band of acoustic states.  These dynamics, which arise from the optical phase-matching conditions, also permit a form of wavevector-resolved phonon spectroscopy that allows us measure the noise spectral density and phonon occupation as a function of acoustic wavevector. This demonstration illustrates a powerful type of phonon control that is not possible in discrete cavity-optomechanical systems. To examine these properties, we perform an additional set of experiments, in which we use strong laser field to cool a group of phonons while measuring the phonon occupancies with a distinct probing field.

As illustrated in Fig. \ref{fig:pumpprobe}a, these experiments use an additional continuous wave pump source of frequency $\omega^{(2)}_{\rm p}$ (distinct from the probe frequency $\omega^{(1)}_{\rm p}$) to cool the phonon field through spontaneous Brillouin scattering. 
Using a much weaker probe field, we simultaneously measure the scattered probe sidebands over a range of probe wavelengths to characterize the modification of the phonon dynamics induced by the strong pump wave.  Due to the optical phase-matching conditions, the center frequencies of the pump and the probe waves determine the set of phonon wavevectors with which each field interacts (for more details, see Supplementary Section 2-3). Thus, pump and probe waves with different center frequencies will address groups of phonons with distinct wavevectors.  In this way, phase-matching permits us to probe the phonon dynamics as a function of acoustic wavevector---simply by tuning the wavelength of the probe field.  This form of wavevector-resolved phonon spectroscopy is performed by coupling both the pump and probe waves into the symmetric mode of the Brillouin-active waveguide. Through spontaneous inter-modal Brillouin scattering, these waves produce two distinct sets of Stokes and anti-Stokes sidebands that propagate in the antisymmetric optical spatial mode. 

To analyze the phonon dynamics, we isolate the spontaneously generated probe sidebands from the fields generated by the pump wave using the frequency selectivity provided by heterodyne detection. By synthesizing a local oscillator (LO) from the probe wave and combining the scattered fields with the LO on a high-speed photo-receiver (Fig. \ref{fig:pumpprobe}a), the Stokes and anti-Stokes light generated by the probe and pump waves produce unique sets of microwave tones that are easily distinguished using a spectrum analyzer. Throughout these experiments, the Stokes (+) and anti-Stokes (-) signals produced by the probe wave appear at respective microwave frequencies $\Omega_{\rm b}\pm\Delta$. By contrast, the scattered light associated with the pump produces microwave signals at vastly different frequencies ($|(\omega^{(2)}_{\rm p}\pm\Omega_{\rm b})-(\omega^{(1)}_{\rm p}+\Delta)|$) that vary with pump-probe detuning. Thus, provided that $|\omega^{(2)}_{\rm p}- \omega^{(1)}_{\rm p}|/(2 \pi)  > 100$ MHz and $|\omega^{(2)}_{\rm p}- (\omega^{(1)}_{\rm p} \pm \Omega_{\rm b})|/(2 \pi)  > 100$ MHz, the Stokes and anti-Stokes sidebands are easily distinguished from the signals produced by the pump wave.





This experimental scheme allows us to perform thermal phonon spectroscopy and observe the cooling effects produced by the pump wave as a function of phonon wavevector.  This is accomplished by keeping the pump wavelength fixed while performing heterodyne spectral analysis over a range of probe wavelengths and pump powers.  The results plotted in Fig. \ref{fig:pumpprobe}b show the dissipation rate difference between Stokes and anti-Stokes spectra and the estimated anti-Stokes phonon occupation as a function of phonon wavevector.  Each data point represents the average change in dissipation rate (and phonon number) that is obtained from a series of spectral measurements as the pump power is increased. Example spectra when the probe wave is (is not) phase-matched to the same group of phonons as the pump wave are plotted in Fig. \ref{fig:pumpprobe}ii (Fig. \ref{fig:pumpprobe}i).   These pump-probe experiments show that spontaneous Brillouin scattering produces cooling over a narrow band of phonon wavevectors, with a wavevector bandwidth inversely proportional to the length of the device (see Supplementary Section 3).  As a result, this form of optomechanical cooling produces a cold window in wavevector space, distinct from the cold frequency windows typically produced by optomechanical resonators \cite{habraken2012continuous,patel2017single}. Thus, this waveguide system gives access to a continuum of phonon modes that can be optomechanically controlled as a function of acoustic wavevector.

\subsection{Cooling dynamics}

Our experimental observations can be explained by a succinct spatio-temporal model, which captures the essential dynamics of continuum optomechanical cooling (see Supplementary Section 1-2).  Starting from a Hamiltonian formalism that describes traveling-wave optomechanical interactions \cite{kharel2016noise,sipe2016hamiltonian}, we calculate the coupled equations of motion for the phonon and anti-Stokes fields.  For the purposes of this derivation, we describe the optical fields as functions of position and the phonon field as a continuous sum over wavevector mode amplitudes. This choice is particularly convenient because the phonon density of states is constant in wavevector space, and the mode occupation is well-defined for each phonon wavevector.

Due to the disparate dissipation rates for the optical ($\gamma$) and acoustic fields ($\Gamma$), we can use separation of time scales (i.e., $\Gamma \ll \gamma$) to greatly simplify the equations of motion, which allows us to eliminate the temporal dynamics of the anti-Stokes field.  In this limit, the anti-Stokes field adiabatically follows the temporal dynamics given by the stochastically (or thermally) driven phonon field.

As derived in Supplementary Section 1, the separation of time scales yields a solution for the effective amplitude of the group of phase-matched phonons that interact with the anti-Stokes field, which is given by
\!
\begin{equation}
\begin{aligned}
\beta(t)&= \int_0^L dz^\prime \int_{-\infty}^{\infty}dq^\prime \int_0^{\infty} dt^\prime \bigg[1-\frac{\Gamma G_{\rm b} P_{\rm p} (L-z^\prime)}{4} t^\prime \bigg] \\
& \times \eta_{q^\prime}(t-t^\prime) e^{-\frac{\Gamma }{2}t^\prime}e^{i(q^\prime-\Delta k_{\rm as})z^\prime}.
\end{aligned}
\label{eq:Bsolfm}
\end{equation}

\noindent Here, $L$ is the length of the interaction region, $\Delta k_{\rm as}$ is the wavevector difference between two optical fields (i.e., $\Delta k_{\rm as} \equiv k_{\rm as} - k_{\rm p}$), $\eta_{q}(t)$ is a Langevin force describing the stochastic thermal driving of the elastic field, and $G_{\rm b}$ is the Brillouin gain coefficient of the Brillouin-active waveguide (with units of $\rm W^{-1} m^{-1}$). 


From Equation \ref{eq:Bsolfm}, we find that the occupation of the phonon field ($\langle n( \Delta k_{\rm as}) \rangle$), relative to the thermal value ($\langle n_{\rm th}( \Delta k_{\rm as}) \rangle$), is given by

\begin{equation}
\begin{aligned}
\frac{\langle n(\Delta k_{\rm as})\rangle}{ \langle n_{\rm th}( \Delta k_{\rm as}) \rangle}&=\frac{\langle \beta^{\dagger}(t)\beta(t) \rangle}{\langle \beta^{\dagger}_{\rm th}(t)\beta_{\rm th}(t) \rangle}\\
&\approx 1- \frac{G_{\rm b} P_{\rm p} L}{4}\\
&\approx \frac{\Gamma}{\Gamma_{\rm as,eff}},
\end{aligned}
\label{eq:occupfm}
\end{equation}

\noindent
in the limit when $(G_{\rm b} P_{\rm p} L/4)^2 \ll 1$. Above, $\Gamma_{\rm as,eff}$ is defined as $\Gamma_{\rm as,eff}= \Gamma\left(1+ G_{\rm b} P_{\rm p} L/4\right)$ (see Supplementary Section 1). From Eq. \ref{eq:occupfm}, we observe that the additional optomechanical damping yields a net reduction in the anti-Stokes phonon occupation.

We next summarize the salient results of our pump-probe analysis, which elucidates the role of phase-matching in the cooling process (presented Supplementary Section 2). The presence of a pump wave yields additional terms in the interaction Hamiltonian, producing a coupled set of equations that describe the phonon dynamics. We find it convenient to solve the coupled phonon dynamics in terms of the fields $\beta^{(1)}(t)$ and $\beta^{(2)}(t)$, representing the phonon bands that interact with the probe and pump waves, respectively.

To simplify our analysis, we assume that the probe wave power is weak ($P_{\rm p}^{(1)}$) and that the pump wave (of power $P_{\rm p}^{(2)}$) modifies the phonon population to first order (i.e., 
$G_{\rm b} P^{(1)}_{\rm p} L/4 \ll 1$ and $(G_{\rm b} P^{(2)}_{\rm p} L/4)^2 \ll 1$), conditions that are well-satisfied for our pump-probe experiments.  In this case, the occupation of the phonon bands interacting with the weak probe wave (relative to the thermal value) is given by

\!
\begin{equation}
\begin{aligned}
\frac{\langle n^{(1)}(\Delta k_{\rm as})\rangle}{ \langle n^{(1)}_{\rm th}( \Delta k_{\rm as}) \rangle}&=\frac{\langle \beta^{(1)\dagger}(t)\beta^{(1)}(t) \rangle}{\langle\beta^{(1)\dagger}_{\rm th}(t)\beta^{(1)}_{\rm th}(t)\rangle}\\
&\approx 1-\frac{G_{\rm b} P^{(2)}_{\rm p}L}{4}\sinc^2\bigg[\frac{(\Delta k^{(2)}_{\rm as}-\Delta k^{(1)}_{\rm as})L}{2}\bigg].
\end{aligned}
\label{eq:occupff}
\end{equation}

Here, superscripts $^{(1)}$ and $^{(2)}$ index the optical fields sourced by the probe and pump waves, respectively. Thus, $\Delta k^{(1)}_{\rm as}$ and $\Delta k^{(2)}_{\rm as}$ are the phase-matching conditions of the probe and pump waves, respectively (i.e., $q_{\rm as}=\Delta k_{\rm as}\equiv k_2(\omega_{\rm p}+\Omega_{\rm b})-k_1(\omega_{\rm p})$, see Supplementary Section 3 for more details). The theoretical trend in Fig. \ref{fig:pumpprobe}b was obtained directly from Eq. \ref{eq:occupff}, which agrees well with our experimental data. In the case of optimal phase-matching (i.e., $\Delta k^{(2)}_{\rm as}=\Delta k^{(1)}_{\rm as}$) and low pump power, this result agrees with the cooling effect produced in the single probe case (see Eq. \ref{eq:occupfm}).

This analysis, in conjunction with our experimental observations, reveals that the optical fields phase-match to a narrow band of phonon wavevectors, with a phase-matching bandwidth (full width at half maximum) determined by the length of the system (i.e., $\Delta q = \Delta k^{(2)}_{\rm as}-\Delta k^{(1)}_{\rm as}=2.78/L$, see Supplementary Section 3). This is particularly striking given that the phonon field is heavily damped in the spatial domain, with an intrinsic decay length of less than 60 $\upmu$m ( $\sim400\times$ shorter than the device length). Nevertheless, due to the disparate velocities of the interacting light and sound fields, optical phase matching plays the dominant role in selecting the phonon wavevectors that participate in the heating and cooling processes.  In this way, the phase-matched optomechanical cooling process can be understood as a form of wavevector-selective reservoir engineering.

An important consequence of these dynamics is that the degree of phonon cooling depends on the length of the device. However, we note that although the cooling efficiency increases with device length, optomechanical cooling occurs over a narrower bandwidth of phonon wavevectors. In addition, the cooling behavior is contingent upon the separation of time scales, which places a limit on the length of the device. If this separation of time scales ($\gamma \gg \Gamma$) is not satisfied, energy transferred from the acoustic field to the optical fields will return to the acoustic field, rather than escaping the system. Thus, longer systems can face a fundamental challenge to achieve Brillouin cooling; the lower limit of the optical dissipation rate, set by the transit time, cannot be smaller than the acoustic dissipation rate.  This sets a practical limitation to the length over which cooling can occur, namely that $L \ll v_{\rm g}/\Gamma$.

\section{Discussion}

In this Letter, we have reported phonon cooling in a continuous optomechanical system, and we show that this form of cooling is possible without an optical cavity or discrete acoustic modes. This demonstration represents an important entry point for new types of optomechanical operations not possible in single-mode (i.e., lumped-element or zero-dimensional) cavity-optomechanical systems \cite{rakich2016quantum}. One intriguing possibility is the use of this type of cooling for reservoir engineering \cite{toth2017dissipative} and nonreciprocal control of thermal phonon transport \cite{seif2018thermal}. Without the frequency constraints imposed by optical cavities, continuous optomechanical systems such as this could prove advantageous for bath engineering over exceptionally large optical and phonon bandwidths \cite{aspelmeyer2014cavity}. Moreover, combining these concepts with novel photonic-phononic emit-receive devices---in which disparate optical waveguides couple through a single traveling-wave phonon field \cite{shin2015control,kittlaus2017integrated}---could unlock a new array of non-local optomechanical phemenona, ranging from mechanically-mediated optical entanglement to non-classical states of light.

In summary, we have demonstrated traveling-wave phonon cooling in a continuous optomechanical system. This cooling is made possible by the large coupling rates and phase-matching-induced symmetry breaking produced by inter-modal Brillouin scattering within an optomechanical silicon waveguide. Through a novel type of thermal phonon spectroscopy, we show that spontaneous Brillouin scattering produces a cold wavevector window in which traveling-wave phonons are optomechanically cooled. This result represents a first step toward wavevector-tunable thermodynamic phonon control, and may open the door to new types of reservoir engineering, nonreciprocal phonon transport, multimode squeezing, and strategies of encoding information in the phononic degrees of freedom in continuous systems.

\subsection{Acknowledgements}
 
We would like to thank Prashanta Kharel, Shai Gertler, and William Renninger for discussions regarding the cooling dynamics and for assistance in developing the spatio-temporal model.

This work was supported through a seedling grant under the direction of Dr. Daniel Green at DARPA MTO and by the Packard Fellowship for Science and Engineering; N.T.O. acknowledges support from the National Science Foundation Graduate Research Fellowship under Grant No. DGE1122492.

\subsection{Author contributions}

N.T.O. and E.A.K. fabricated the silicon waveguide devices. \:N.T.O. and E.A.K. performed experiments with guidance from P.T.R..\: R.O.B., N.T.O., and P.T.R. developed analytic models of the Brillouin cooling dynamics. \:  N.T.O. performed data analysis with assistance from E.A.K. and R.O.B.. All authors made contributions to the writing of the manuscript.

\bibliography{cites}

\newpage\null\thispagestyle{empty}\newpage


\section*{Supplementary material}

# Supplementary Information: Optomechanical cooling in a continuous system

Nils T. Otterstrom,[1] Ryan O. Behunin,[2] Eric A. Kittlaus,[1] and Peter T. Rakich[1]

[1]*Department of Applied Physics, Yale University, New Haven, CT 06520 USA.*

[2]*Department of Physics and Astronomy, Northern Arizona University, Flagstaff, AZ 86001 USA.*

(Dated: May 5, 2018)

## Contents

# 1. CONTINUUM OPTOMECHANICAL COOLING DYNAMICS

This section outlines a derivation for the effect of spontaneous inter-modal Brillouin scattering on the anti-Stokes phonon lifetime and population. The intrinsic decoupling between Stokes and anti-Stokes processes allows us to derive the dynamics of each process separately. Starting from the formalism developed in Ref. [1], the Hamiltonian for the spontaneous Brillouin interaction can be expressed as

$$
\begin{aligned}
H = & \hbar \int_{-\infty}^{\infty} dz' A_{\mathrm{p}}^{\dagger}(z',t)\omega_1(\hat{k}_{z'})A_{\mathrm{p}}(z',t) \\
& + \hbar \int_{-\infty}^{\infty} dz' A_{\mathrm{as}}^{\dagger}(z',t)\omega_2(\hat{k}_{z'})A_{\mathrm{as}}(z',t) \\
& + \hbar \int_{-\infty}^{\infty} dq' \Omega_{q'} b_{q'}^{\dagger} b_{q'} \\
& + \frac{\hbar}{\sqrt{2\pi}} \int_{-\infty}^{\infty} dq' \int_{0}^{L} dz' g_0 A_{\mathrm{as}}^{\dagger}(z',t)A_{\mathrm{p}}(z',t)b_{q'}e^{i(q'-\Delta k_{\mathrm{as}})z'} \\
& + \frac{\hbar}{\sqrt{2\pi}} \int_{-\infty}^{\infty} dq' \int_{0}^{L} dz' g_0^* A_{\mathrm{as}}(z',t)A_{p}^{\dagger}(z',t)b_{q'}^{\dagger}e^{-i(q'-\Delta k_{\mathrm{as}})z'}.
\end{aligned}
\tag{1}
$$

Here, the wavevector difference between pump and anti-Stokes fields is defined by $\Delta k_{\mathrm{as}} \equiv k_2(\omega_{\mathrm{as}}) - k_1(\omega_{\mathrm{p}})$, $g_0$ is the Brillouin coupling rate given by the acoustic-optic overlap and photo-elastic tensor, and the Taylor expansion of the symmetric and antisymmetric optical mode dispersion relations in the $z$ basis are given by $\omega_i(\hat{k}_z) \equiv \sum_{n=0}^{\infty} \frac{1}{n!} \frac{\partial^n \omega_i}{\partial k^n}(-i\frac{\partial}{\partial z})^n$.

From the Heisenberg equations of motion, we calculate the spatio-temporal dynamics of the continuous phonon and anti-Stokes fields. Using the fact that $[A_{\mathrm{as}}(z',t), A_{\mathrm{as}}^{\dagger}(z,t)] = \delta(z-z')$ and $[b_{q'}(t), b_q^{\dagger}(t)] = \delta(q-q')$ [1], these equations are computed to be

$$
\begin{aligned}
\dot{b}_q(t) = & -i\Omega_q b_q(t) - \frac{\Gamma}{2}b_q(t) + \eta_q(t) - \frac{ig_0^* A_{\mathrm{p}}^{\dagger}(t)}{\sqrt{2\pi}} \int_{0}^{L} dz' A_{\mathrm{as}}(z',t)e^{-i(q-\Delta k_{\mathrm{as}})z'} \\
\dot{A}_{\mathrm{as}}(z,t) = & -v_{\mathrm{g}}\frac{\partial}{\partial z}A_{\mathrm{as}}(z,t) - i\omega_{\mathrm{as}}A_{\mathrm{as}}(z,t) - \frac{\gamma}{2}A_{\mathrm{as}}(z,t) - \frac{ig_0 A_{\mathrm{p}}(t)}{\sqrt{2\pi}} \int_{-\infty}^{\infty} dq' b_{q'}(t)e^{i(q'-\Delta k_{\mathrm{as}})z}.
\end{aligned}
\tag{2}
$$

Here, we pause briefly to summarize the steps and assumptions involved in this derivation. We assume the that the probe wave ($A_{\mathrm{p}}$) remains undepleted through the spontaneous process (i.e., $A_{\mathrm{p}}(z,t) = A_{\mathrm{p}}(t)$), which is very well satisfied in the present system. To include the effects of dissipation, we introduce the phonon decay rate, $\Gamma$, and an associated Langevin stochastic driving term $\eta_q(t)$, which maintains the system in thermal equilibrium in the absence of optical forcing; $\eta_q(t)$ has a two-time correlation function given by $\langle \eta_{q'}^{\dagger}(t')\eta_q(t)\rangle = Q\delta(t-t')\delta(q-q')$, where $Q$ is given by the thermal equilibrium properties of the system and is consistent with the fluctuation dissipation theorem; in this case, $Q = n_{\mathrm{th}}\Gamma$, where $n_{\mathrm{th}}$ is given by the Planck distribution ($n_{\mathrm{th}} = [\exp(\hbar\Omega_b/(k_{\mathrm{b}}T)) - 1]^{-1}$). We treat the optical fields classically and ignore the Langevin driving force associated with the optical anti-Stokes dissipation rate ($\gamma$), since the thermal occupation of the anti-Stokes field is, for our purposes, negligible (i.e., $n_{\mathrm{as}} \cong 10^{-14}$). For the anti-Stokes field, we limit the Taylor expansion of the optical dispersion relation to first order and ignore higher order dispersion.

Next, we move into the rotating frame to formally solve the equation of motion for the anti-Stokes field. By introducing $A_{\mathrm{as}}(z,t) = \bar{A}_{\mathrm{as}}(z,t)e^{-i\omega t}$, $A_{\mathrm{p}}(z,t) = \bar{A}_{\mathrm{p}}(z,t)e^{-i\omega_{\mathrm{p}}t}$, and $b_q(t) = \bar{b}_q(t)e^{-i\Omega t}$ (where $\Omega = \omega - \omega_{\mathrm{p}}$), the equation of motion for the anti-Stokes wave in the rotating frame becomes

$$
\dot{\bar{A}}_{\mathrm{as}}(z,t) = -v_{\mathrm{g}}\frac{\partial}{\partial z}\bar{A}_{\mathrm{as}}(z,t) + [-i(\omega_{\mathrm{as}} - \omega) - \frac{\gamma}{2}]\bar{A}_{\mathrm{as}}(z,t) - \frac{ig_0\bar{A}_{\mathrm{p}}(t)}{\sqrt{2\pi}} \int_{-\infty}^{\infty} dq' \bar{b}_{q'}(t)e^{i(q'-\Delta k_{\mathrm{as}})z}.
\tag{3}
$$

Since, in this system, $\gamma \gg \Gamma$ and $\gamma \gg |\omega_{\mathrm{as}} - \omega|$, we can adiabatically eliminate the temporal dynamics of the anti-Stokes field (i.e., $\dot{\bar{A}}_{\mathrm{as}} = 0$). In other words, because the anti-Stokes field is heavily damped in time, the temporal dynamics of the anti-Stokes field adiabatically follow those of the phonon field. In this limit, the anti-Stokes mode envelope amplitude is given by

$$
\bar{A}_{\mathrm{as}}(z,t) = \frac{-ig_0\bar{A}_{\mathrm{p}}(t)}{v_{\mathrm{g}}\sqrt{2\pi}} \int_{0}^{z} dz' \int_{-\infty}^{\infty} dq' \bar{b}_{q'}e^{i(q'-\Delta k_{\mathrm{as}})z' - \frac{\gamma}{2v_{\mathrm{g},2}}(z-z')}.
\tag{4}
$$

Since $2v_{\mathrm{g}} \gg \gamma L$, the factor $\exp\{-\frac{\gamma}{2v_{\mathrm{g},2}}(z-z')\}$ within the integrand is, to a good approximation, unity. As a result, we can insert this solution into the equation of motion for the phonon field to find

$$
\dot{\bar{b}}_q = -i(\Omega_q - \Omega)\bar{b}_q - \frac{\Gamma}{2}\bar{b}_q + \eta_q(t) - \frac{|g_0|^2|\bar{A}_p|^2}{v_{\mathrm{g},2}(2\pi)} \int_{0}^{L} dz' \int_{0}^{z'} dz'' \int_{-\infty}^{\infty} dq' \bar{b}_{q'}e^{i(q'-\Delta k_{\mathrm{as}})z'' - i(q-\Delta k_{\mathrm{as}})z'}
\tag{5}
$$

We proceed by first performing the integrals over space. The resulting integrand can then be separated into real and imaginary parts. We are interested in the case where $q = \Delta k_{as}$ and $\Omega = \Omega_{\Delta k_{as}}$, or when the phonon wavevector is defined by optimal phase-matching with the optical fields. These conditions yield

$$\dot{\bar{b}}_{\Delta k_{as}} = -\frac{\Gamma}{2}\bar{b}_{\Delta k_{as}} + \eta_{\Delta k_{as}}(t) - \frac{|g_0|^2|A_p|^2}{v_{g,2}(2\pi)}\int_{-\infty}^{\infty}dq'\bar{b}_{q'}\big[\frac{1-\cos\left[L(\Delta k_{as}-q')\right]}{(\Delta k_{as}-q')^2} + i\frac{L(q'-\Delta k_{as})+\sin\left[L(\Delta k_{as}-q')\right]}{(\Delta k_{as}-q')^2}\big].\tag{6}$$

The real and imaginary coefficients multiplying $\bar{b}_{q'}$ in the integrand are sharply peaked even and odd functions (about $q' = \Delta k_{as}$), respectively. We also note that $\bar{b}_{q'}$ is nearly symmetrically distributed about $q' = \Delta k_{as}$ due to the thermal distribution. These conditions yield an approximate equation of motion for the phonon field, given by

$$\dot{b}_{\Delta k_{as}} = -i\Omega_{\Delta k_{as}}b_{\Delta k_{as}} - \frac{\Gamma_{as,eff}}{2}b_{\Delta k_{as}} + \eta_{\Delta k_{as}}(t),\tag{7}$$

where $\Gamma_{as,eff} = \Gamma\left(1 + G_b P_p L/4\right)$.

Here, $P_p$ is the probe power and $G_b$ is the Brillouin gain coefficient (with units $W^{-1}m^{-1}$) defined by $G_b = 4|g_0|^2|A_p|^2/(P_p\Gamma v_{g,2})$. Equation 7 reveals that the presence of the strong pump modifies the equation of motion for the phonon through an additional damping term. The anti-Stokes process annihilates a phonon and a pump photon to generate an anti-Stokes (or blue-shifted) photon. Hence, the anti-Stokes process reduces the lifetime of the phonon, evident as a correction to the damping rate.

We now return to Eq. 5, in order to make a transformation that simplifies our analysis and provides a direct connection to our measurements. We begin by defining

$$\beta(z,t) \equiv \int_0^z dz'\int_{-\infty}^{\infty}dq'\bar{b}_{q'}(t)e^{i(q'-\Delta k_{as})z'}$$
$$\xi(z,t) \equiv \int_0^z dz'\int_{-\infty}^{\infty}dq'\eta_{q'}(t)e^{i(q'-\Delta k_{as})z'}.\tag{8}$$

Here, $\beta(z,t)$ represents the amplitude of the band of phase-matched phonons that interact with the probe wave (see Eq. 4), and $\xi(z,t)$ is the Langevin force under this transformation. Re-expressing Eq. 5 in terms of $\beta(z,t)$ and $\xi(z,t)$ yields

$$\dot{\beta}(z,t) = -\frac{\Gamma}{2}\beta(z,t) + \xi(z,t) - \frac{|g_0|^2|A_p|^2}{v_g}\int_0^z dz'\beta(z',t).\tag{9}$$

We proceed by performing a Fourier transform in time and taking a spatial derivative, which yields

$$\frac{dB[z,\omega]}{dz} = -\frac{\chi G_b P_p \Gamma}{4}B[z,\omega] + \chi\int_{-\infty}^{\infty}dq'\tilde{\eta}_{q'}e^{i(q'-\Delta k_{as}^{(1)})z}.\tag{10}$$

Here, we have defined $\chi \equiv 1/(-i\omega+\Gamma/2)$, $B[z,\omega]$ is the Fourier transform of $\beta(z,t)$, and $\tilde{\eta}_{q'}$ is the Fourier transformed Langevin force (i.e., $\tilde{\eta}_{q'}[\omega] = \mathscr{F}[\eta_{q'}] \equiv 1/(2\pi)\int_{-\infty}^{\infty}\eta_{q'}(t)e^{-i\omega t}$). We have also used the fact that $|g_0|^2|A_p|^2/v_g = G_b P_p \Gamma/4$ [1].

Solving Eq. 10 yields

$$B[z,\omega] = \int_0^z dz'\int_{-\infty}^{\infty}dq'\chi\tilde{\eta}_{q'}e^{-\frac{\chi\Gamma P_p(z-z')}{4}}e^{i(q'-\Delta k_{as})z'}.\tag{11}$$

To simplify the inverse Fourier transform, we Taylor expand $e^{-\frac{\chi\Gamma P_p(z-z')}{4}}$ to first order, which is a good approximation in view of the probe power, Brillouin gain coefficient, and length of this system. We then perform an inverse Fourier transform via the convolution theorem, yielding the solution for $\beta(z,t)$, which is given by

$$\beta(z,t) = \int_0^z dz'\int_{-\infty}^{\infty}dq'\int_0^{\infty}dt'\Big[1 - \frac{\Gamma G_b P_p(z-z')}{4}t'\Big]\eta_{q'}(t-t')e^{-\frac{\Gamma}{2}t'}e^{i(q'-\Delta k_{as})z'}.\tag{12}$$

In the absence of the probe field (i.e., $P_p = 0$), the solution for the thermal phonon field (i.e., $\beta_{th}(z,t)$) is given by

$$\beta_{th}(z,t) = \int_0^z dz'\int_{-\infty}^{\infty}dq'\int_0^{\infty}dt'\eta_{q'}(t-t')e^{-\frac{\Gamma}{2}t'}e^{i(q'-\Delta k_{as})z'}.\tag{13}$$

As a result, the modified occupation of the band of phonons interacting with the probe wave ($\langle n(\Delta k_{\mathrm{as}})\rangle$) relative the thermal occupation ($\langle n_{\mathrm{th}}(\Delta k_{\mathrm{as}})\rangle$) is given by

$$
\begin{aligned}
\frac{\langle n(\Delta k_{\mathrm{as}})\rangle}{\langle n_{\mathrm{th}}(\Delta k_{\mathrm{as}})\rangle} &= \frac{\langle \beta^\dagger(L,t)\beta(L,t)\rangle}{\langle \beta_{\mathrm{th}}^\dagger(L,t)\beta_{\mathrm{th}}(L,t)\rangle} \\
&= 1 - \frac{G_{\mathrm{b}}P_{\mathrm{p}}L}{4} + \frac{G_{\mathrm{b}}^2 P_{\mathrm{p}}^2 L^2}{24} \\
&\approx 1 - \frac{G_{\mathrm{b}}P_{\mathrm{p}}L}{4} \\
&\approx \frac{\Gamma}{\Gamma_{\mathrm{as,eff}}},
\end{aligned}
\tag{14}
$$

where the approximations assume that $(G_{\mathrm{b}}P_{\mathrm{p}}L/4)^2 \ll 1$. From Eq. 14, we observe that the additional damping produced by spontaneous anti-Stokes scattering pushes the phonon field out of thermal equilibrium, yielding a reduction in the phonon occupation.

Following a similar derivation one can show that the effective dissipation rate for Stokes phonons is $\Gamma_{\mathrm{s,eff}} = \Gamma\left(1 - G_{\mathrm{b}}P_{\mathrm{p}}L/4\right)$, and the phonon occupation is

$$
\begin{aligned}
\frac{\langle n(\Delta k_{\mathrm{as}})\rangle}{\langle n_{\mathrm{th}}(\Delta k_{\mathrm{as}})\rangle} &= 1 + \frac{G_{\mathrm{b}}P_{\mathrm{p}}L}{4} + \frac{G_{\mathrm{b}}^2 P_{\mathrm{p}}^2 L^2}{24} \\
&\approx 1 + \frac{G_{\mathrm{b}}P_{\mathrm{p}}L}{4} \\
&\approx \frac{\Gamma}{\Gamma_{\mathrm{s,eff}}}
\end{aligned}
\tag{15}
$$

Thus, the spontaneous Stokes process reduces the damping and increases the thermal occupation of the phonon field.

## 2. PUMP-PROBE COOLING THEORY

In this section, we derive the dynamics of our pump-probe experiments (Fig. 3) and show how phase-matching allows us to perform wavevector-resolved measurements of the phonon occupation. In these measurements, we use a strong pump wave (of amplitude $A_{\mathrm{p}}^{(2)}$) to modify the phonon dynamics and a weak probe wave (of amplitude $A_{\mathrm{p}}^{(1)}$) to measure the phonon occupation. In this case, the interaction Hamiltonian of the system is given by

$$
\begin{aligned}
H_{\mathrm{int}} = \frac{\hbar}{\sqrt{2\pi}} \times & \\
\Bigg[ &\int_{-\infty}^{\infty} dq' \int_0^L dz' g_0 A_{\mathrm{as}}^{(1)\dagger}(z',t) A_{\mathrm{p}}^{(1)}(z',t) b_{q'} e^{i(q'-\Delta k_{\mathrm{as}}^{(1)})z'} \\
&+ \int_{-\infty}^{\infty} dq' \int_0^L dz' g_0^* A_{\mathrm{as}}^{(1)}(z',t) A_{p}^{(1)\dagger}(z',t) b_{q'}^\dagger e^{-i(q'-\Delta k_{\mathrm{as}}^{(1)})z'} \\
&+ \int_{-\infty}^{\infty} dq' \int_0^L dz' g_0 A_{\mathrm{as}}^{(2)\dagger}(z',t) A_{\mathrm{p}}^{(2)}(z',t) b_{q'} e^{i(q'-\Delta k_{\mathrm{as}}^{(2)})z'} \\
&+ \int_{-\infty}^{\infty} dq' \int_0^L dz' g_0^* A_{\mathrm{as}}^{(2)}(z',t) A_{p}^{(2)\dagger}(z',t) b_{q'}^\dagger e^{-i(q'-\Delta k_{\mathrm{as}}^{(2)})z'} \Bigg].
\end{aligned}
\tag{16}
$$

Here, superscripts $^{(1)}$ and $^{(2)}$ index the optical fields sourced by the probe and pump waves, respectively. Here, we note that the phase-matching conditions $\Delta k_{\mathrm{as}}^{(1)}$ and $\Delta k_{\mathrm{as}}^{(2)}$ are given simply by the wavelengths of the probe and pump waves, respectively, which in general are distinct (for further details see Section 3). We next compute and simplify the equations of motion following the same approach used to develop Eqs. 1-5. This leads to the following equation for the phonon field of wavevector $q$

$$\dot{\bar{b}}_q = -i(\Omega_q - \Omega)\bar{b}_q - \frac{\Gamma}{2}\bar{b}_q - \eta_q(t)$$
$$- \frac{|g_0|^2|\bar{A}_p^{(1)}|^2}{v_g(2\pi)}\int_0^L dz' \int_0^{z'} dz'' \int_{-\infty}^{\infty} dq' \bar{b}_{q'} e^{i(q'-\Delta k_{as}^{(1)})z'' - i(q-\Delta k_{as}^{(1)})z'}$$
$$- \frac{|g_0|^2|\bar{A}_p^{(2)}|^2}{v_g(2\pi)}\int_0^L dz' \int_0^{z'} dz'' \int_{-\infty}^{\infty} dq' \bar{b}_{q'} e^{i(q'-\Delta k_{as}^{(2)})z'' - i(q-\Delta k_{as}^{(2)})z'}. \tag{17}$$

From Eq. 17, it is evident that both the probe and pump waves can contribute to Brillouin cooling as long as they satisfy phase matching. This produces a system of equations describing the coupled dynamics of the the phonon bands interacting with the probe and pump waves (i.e., setting $q = \Delta k_{as}^{(1)}$ and $q = \Delta k_{as}^{(2)}$, respectively).

We now transform Eq. 17 using the same procedure used to obtain Eq. 9. We begin by defining

$$\beta^{(m)}(z,t) \equiv \int_0^z dz' \int_{-\infty}^{\infty} dq' \bar{b}_{q'}(t) e^{i(q'-\Delta k_{as}^{(m)})z'}, \tag{18}$$

where $m = (1,2)$. From Eq. 18, it can be seen that $\beta^{(1)}(z,t)$ and $\beta^{(2)}(z,t)$ represent the narrow bands of phonons that interact with the probe and pump waves, respectively, as dictated by phase-matching. Rewriting Eq. 17 in terms of these new spatially dependent functions yields the following coupled integral equations given by

$$\dot{\beta}^{(1)}(z,t) = -\frac{\Gamma}{2}\beta^{(1)}(z,t) + \xi^{(1)}(z,t)$$
$$- \frac{|g_0|^2|A_p^{(1)}|^2}{v_g}\int_0^z dz' \beta^{(1)}(z',t) - \frac{|g_0|^2|A_p^{(2)}|^2}{v_{g,2}}\int_0^z dz' \beta^{(2)}(z',t) e^{i(\Delta k_{as}^{(2)} - \Delta k_{as}^{(1)})z'}$$
$$\dot{\beta}^{(2)}(z,t) = -\frac{\Gamma}{2}\beta^{(2)}(z,t) + \xi^{(2)}(z,t)$$
$$- \frac{|g_0|^2|A_p^{(1)}|^2}{v_g}\int_0^z dz' \beta^{(2)}(z',t) - \frac{|g_0|^2|A_p^{(1)}|^2}{v_{g,2}}\int_0^z dz' \beta^{(1)}(z',t) e^{i(\Delta k_{as}^{(1)} - \Delta k_{as}^{(2)})z'}. \tag{19}$$

Here, $\xi^{(1,2)}(z,t)$ is the Langevin force describing the thermal fluctuations of these phonon bands, defined by $\xi^{(1,2)}(z,t) \equiv \int_0^z dz' \int_{-\infty}^{\infty} dq' \eta_{q'}(t) e^{i(q'-\Delta k_{as}^{(1,2)})z'}$. We next note that $|g_0|^2|A_p^{(1,2)}|^2/v_g = G_b P_p^{(1,2)}\Gamma/4$, where $P_p^{(1,2)}$ is the power of the probe and pump waves, respectively. Rewriting Eq. 19 in terms of these constants and taking the Fourier transform in time yields

$$-i\omega B^{(1)}[z,\omega] = -\frac{\Gamma}{2}B^{(1)}[z,\omega] + \tilde{\xi}^{(1)}[z,\omega]$$
$$- \frac{G_b P_p^{(1)}\Gamma}{4}\int_0^z dz' B^{(1)}[z',\omega] - \frac{G_b P_p^{(2)}\Gamma}{4}\int_0^z dz' B^{(2)}[z',\omega] e^{i(\Delta k_{as}^{(2)} - \Delta k_{as}^{(1)})z'}$$
$$-i\omega B^{(2)}[z,\omega] = -\frac{\Gamma}{2}B^{(2)}[z,\omega] + \tilde{\xi}^{(2)}[z,\omega]$$
$$- \frac{G_b P_p^{(2)}\Gamma}{4}\int_0^z dz' B^{(2)}[z',\omega] - \frac{G_b P_p^{(1)}\Gamma}{4}\int_0^z dz' B^{(1)}[z',\omega] e^{i(\Delta k_{as}^{(1)} - \Delta k_{as}^{(2)})z'}. \tag{20}$$

Here, $B^{(1,2)}[z,\omega]$ and $\tilde{\xi}^{(1,2)}[z,\omega]$ are the Fourier-transformed phonon fields and Langevin force, respectively. Taking the spatial derivative of these equations yields

$$\frac{dB^{(1)}[z,\omega]}{dz} = -\frac{\chi G_b P_p^{(1)}\Gamma}{4}B^{(1)}[z,\omega] - \frac{\chi G_b P_p^{(2)}\Gamma}{4}B^{(2)}[z,\omega]e^{i(\Delta k_{as}^{(2)} - \Delta k_{as}^{(1)})z} + \chi \int_{-\infty}^{\infty} dq' \tilde{\eta}_{q'} e^{i(q'-\Delta k_{as}^{(1)})z}$$
$$\frac{dB^{(2)}[z,\omega]}{dz} = -\frac{\chi G_b P_p^{(2)}\Gamma}{4}B^{(2)}[z,\omega] - \frac{\chi G_b P_p^{(1)}\Gamma}{4}B^{(1)}[z,\omega]e^{i(\Delta k_{as}^{(1)} - \Delta k_{as}^{(2)})z} + \chi \int_{-\infty}^{\infty} dq' \tilde{\eta}_{q'} e^{i(q'-\Delta k_{as}^{(2)})z}. \tag{21}$$

Here, $\chi \equiv 1/(-i\omega + \Gamma/2)$ and $\tilde{\eta}_{q'}$ is the Fourier-transformed Langevin force (i.e., $\tilde{\eta}_{q'}[\omega] = \mathscr{F}[\eta_{q'}] \equiv 1/(2\pi)\int_{-\infty}^{\infty} \eta_{q'}(t)e^{-i\omega t}$). We solve Eq. 21 perturbatively in a manner that is consistent with our experimental parameters. Here we assume that the probe wave is weak and that the strong pump wave modifies the phonon dynamics to first order. This analysis requires that $G_b P_p^{(1)}L/4 \ll 1$ and $(G_b P_p^{(2)}L/4)^2 \ll 1$, which is well-satisfied in our pump-probe experiments. Under this approximation, Eq. 21 simplifies to

$$B^{(1)}[z,\omega] = \int_{-\infty}^{\infty} dq' \int_0^z dz' \chi \tilde{\eta}_{q'} e^{i(q' - \Delta k_{as}^{(1)})z'} - \frac{G_b P_p^{(2)} \Gamma}{4} \int_0^L dz' B^{(2)}[z',\omega] e^{i(\Delta k_{as}^{(2)} - \Delta k_{as}^{(1)})z'}$$

$$B^{(2)}[z,\omega] = \int_{-\infty}^{\infty} dq' \int_0^z dz' \chi \tilde{\eta}_{q'} e^{i(q' - \Delta k_{as}^{(2)})z'}$$

(22)

Substituting $B^{(2)}[z,\omega]$ into the equation for $B^{(1)}[z,\omega]$ and performing the inverse Fourier transform yields

$$\beta^{(1)}(z,t) = \int_{-\infty}^{\infty} dq' \int_0^z dz' \int_0^{\infty} dt' \eta_{q'}(t-t') e^{-\frac{\Gamma}{2}t'} e^{i(q' - \Delta k_{as}^{(1)})z'}$$
$$- \frac{G_b P_p^{(2)} \Gamma}{4} \int_{-\infty}^{\infty} dq' \int_0^z dz' \int_0^{z'} dz'' \int_0^{\infty} dt' \eta_{q'}(t-t') t' e^{-\frac{\Gamma}{2}t'} e^{i(q' - \Delta k_{as}^{(2)})z''} e^{i(\Delta k_{as}^{(2)} - \Delta k_{as}^{(1)})z'}.$$

(23)

From Eq. 23, we compute the occupation of the group of phonons interacting with the probe wave relative to the equilibrium value. This computation yields

$$\frac{\langle n^{(1)}(\Delta k_{as})\rangle}{\langle n_{th}^{(1)}(\Delta k_{as})\rangle} = \frac{\langle \beta^{(1)\dagger}(L,t)\beta^{(1)}(L,t)\rangle}{\langle \beta_{th}^{(1)\dagger}(L,t)\beta_{th}^{(1)}(L,t)\rangle}$$
$$= 1 - \frac{G_b P_p^{(2)} L}{4} \text{sinc}^2 \left[ \frac{(\Delta k_{as}^{(2)} - \Delta k_{as}^{(1)})L}{2} \right]$$
$$+ O\left( \left[ \frac{G_b P_p^{(2)} L}{4} \right]^2 \right)$$
$$\approx 1 - \frac{G_b P_p^{(2)} L}{4} \text{sinc}^2 \left[ \frac{(\Delta k_{as}^{(2)} - \Delta k_{as}^{(1)})L}{2} \right].$$

(24)

This is the central result of this section. Thus, if the necessary conditions are satisfied (i.e., $G_b P_p^{(1)} L/4 \ll 1$ and $(G_b P_p^{(2)} L/4)^2 \ll 1$), we can, to an excellent approximation, use the anti-Stokes light generated from the probe wave to measure the phonon occupation and effective temperature of the band of phonons that are phase-matched to the probe wave. It is evident that Eq. 24 reaches a minimum when the pump wave is optimally phase-matched to the probe wave (i.e., $\Delta k_{as}^{(2)} = \Delta k_{as}^{(1)}$), which occurs when the probe and pump wavelengths are equal $\lambda_p^{(1)} = \lambda_p^{(2)}$ (see Section 3). In this case, the phonon occupation is

$$\frac{\langle n^{(1)}(\Delta k_{as})\rangle}{\langle n_{th}^{(1)}(\Delta k_{as})\rangle} \approx 1 - \frac{G_b P_p^{(2)} L}{4},$$

(25)

which, as a cross-check, agrees with the single-probe theory (see Eq. 14).

## 3. PHASE-MATCHING BANDWIDTH

Incorporating the known parameters of our system, we next calculate the phase-matching bandwidth of phonon wavevectors that are appreciably cooled through spontaneous inter-modal Brillouin scattering. We begin by stating the phase-matching condition required by the anti-Stokes scattering process of the probe wave, namely that $q_{as}^{(1)}(\Omega_b) = \Delta k_{as}^{(1)} \equiv k_2(\omega_p^{(1)} + \Omega_b) - k_1(\omega_p^{(1)})$, where $q_{as}^{(1)}(\Omega_b)$ is the wavevector of the phonon field interacting with the probe wave. Thus, the wavevector of the phonon field is dictated by the frequencies and propagation constants two optical fields. Now, if we consider the phonon wavevector that interacts with the pump frequency ($q_{as}^{(2)}(\Omega_b)$), such that $\omega_p^{(2)} = \omega_p^{(1)} + \Delta\omega$, we find that the difference in the two phonon wavevectors is given by

$$q_{as}^{(2)}(\Omega_b) - q_{as}^{(1)}(\Omega_b) = \Delta k_{as}^{(2)} - \Delta k_{as}^{(1)}$$
$$= k_2(\omega_p^{(2)} + \Omega_b) - k_1(\omega_p^{(2)})$$
$$- [k_2(\omega_p^{(1)} + \Omega_b) - k_1(\omega_p^{(1)})].$$

(26)

Ignoring higher order dispersion, we can simplify this expression in terms of the group velocities of the two interacting optical spatial modes ($v_{g,1}$ and $v_{g,2}$ for the symmetric and antisymmetric spatial mode, respectively). This simplification yields an expression given by

$$\Delta k_{\mathrm{as}}^{(2)} - \Delta k_{\mathrm{as}}^{(1)} = \left[\frac{1}{v_{g,2}} - \frac{1}{v_{g,1}}\right]\Delta\omega.$$

(27)

Here, $\Delta\omega = \omega_{\mathrm{p}}^{(2)} - \omega_{\mathrm{p}}^{(1)}$. The phase-matching bandwidth is given by Eq. 24, which is shown in Fig. 3. As a result, the full width at half maximum (FWHM) is $(\Delta k_{\mathrm{as}}^{(2)} - \Delta k_{\mathrm{as}}^{(1)}) = 2.78/L$. Thus, in $k$-space, the phase-matching bandwidth is simply determined by the length of the waveguide. As a result, the FWHM of the phase-matching bandwidth (in $\Delta f = \Delta\omega/2\pi$) becomes

$$\mathrm{FWHM} = \frac{2.78}{\pi L}\left[\frac{1}{\left|\frac{1}{v_{g,1}} - \frac{1}{v_{g,2}}\right|}\right].$$

(28)

Inserting the parameters of this silicon system (detailed in Methods), we find that the phase-matching bandwidth is 91.4 GHz (0.72 nm). This calculation finds good qualitative agreement with the phase-matching bandwidth exhibited by the pump probe experiments (see Fig. 3). Thus, by shifting the probe wavelength by several nm, we can cool an entirely different group of phonons. In this way, the conditions set by phase-matching permit cooling of continuously accessible groups of phonons in a wavevector-selective fashion.

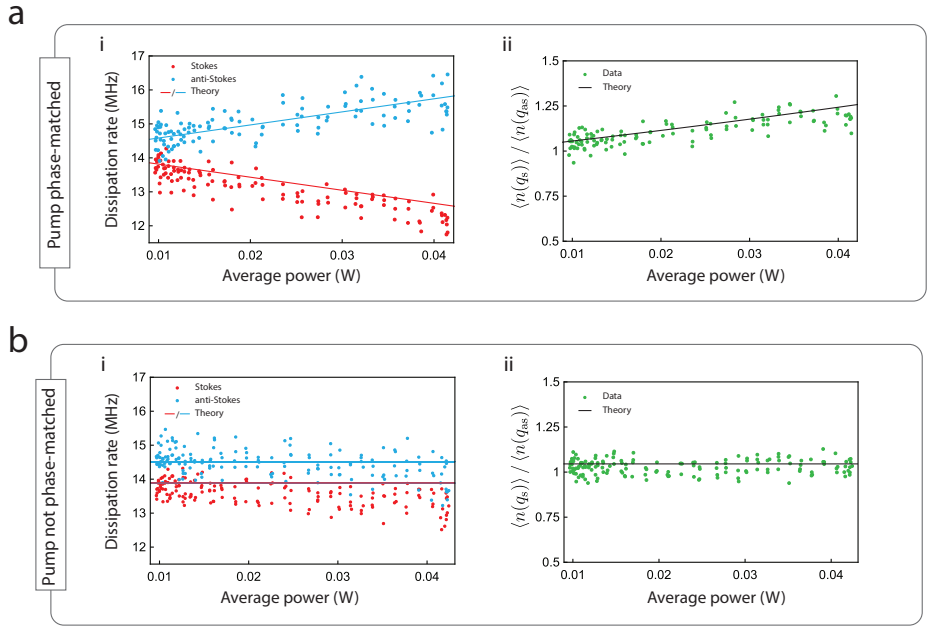

FIG. S1: Pump-probe spontaneous Brillouin scattering measurements. Panel (a)i-ii plots the salient effects of a strong pump wave on the probe spectra when the pump is within the phase matching bandwidth ($\lambda_{\mathrm{p}}^{(2)} = 1535.6$ nm, $\lambda_{\mathrm{p}}^{(1)} = 1535.53$ nm). When this phase-matching condition is satisfied (i.e., the pump wave is interacting with the same band of phonon wavevectors), we observe that the anti-Stokes (Stokes) probe spectra experiences linewidth broadening (narrowing) as the pump power is increased (see panel (ai)), revealing that the pump wave coherently modifies the lifetimes of the Stokes and anti-Stokes phonons. Panel (aii) plots the ratio of Stokes to anti-Stokes power (proportional to the ratio of phonon occupations) as a function of total on-chip power and the theoretical trend from the expected change in Stokes and anti-Stokes phonon occupations. Panel (b)i-ii plots the spectral widths and relative powers of the Stokes and anti-Stokes light (along with associated theoretical trends) when the pump wave is not phase-matched to the phonons interacting with the probe wave ($\lambda_{\mathrm{p}}^{(2)} = 1537.3$ nm, $\lambda_{\mathrm{p}}^{(1)} = 1535.53$ nm). We observe that the spectral widths and ratio of Stokes to anti-Stokes powers remains constant as a function of pump power, which is consistent with the theoretical trend (black). Note that the constant dissipation rate asymmetry is due to the presence of the probe wave.

## 4. ADDITIONAL MEASUREMENTS

In this section, we present additional pump-probe measurements, which are shown in Fig. S1. These pump-probe experiments are performed using the setup diagrammed in Fig. 3a, with a fixed on-chip probe power of 8 mW and

a variable pump power. We first perform a series of spontaneous measurements when the pump is within the phase bandwidth ($\lambda_p^{(2)} = 1535.6$ nm, $\lambda_p^{(1)} = 1535.53$ nm). The salient properties of these measurements are plotted as a function of total on-chip power in Fig. S1ai-ii. Figure S1ai plots the dissipation rates of the Stokes and anti-Stokes spectra, while Fig. S1aii plots the ratio of Stokes to anti-Stokes phonon occupations. From these data, it is evident that as the pump power increases, the anti-Stokes phonons experience a shorter lifetime and consequently a reduction in population, while the Stokes phonon lifetime and population are enhanced. By contrast, for the case when the pump is outside the phase-matching bandwidth, the linewidth and occupation asymmetry between the Stokes and anti-Stokes spectra remain unchanged as the pump power increases, as seen in Fig. S1bi-ii.

These results (1) demonstrate that the dissipation rates can be used to faithfully estimate the phonon occupations (as evident in Eq. 14) and (2) rule out the possibility that these effects are due to optical-only gain filtering (i.e., spectral narrowing or broadening associated with optical nonlinear gain or loss, respectively).

## 5. DEVICE FABRICATION

The Brillouin-active waveguide devices are fabricated from a single-crystal silicon-on-insulator (SOI) wafer with a 3-μm SiO$_2$ layer beneath a 215-nm layer of crystalline silicon (see Refs. [2–4] for ancillary details). Optical ridge waveguides are patterned using electron beam lithography on negative hydrogen silsesquioxane photoresist (HSQ) photoresist. After development in Microposit$^{TM}$ MF$^{TM}$-312, a reactive ion chlorine etch (RIE) is used to remove 80 nm of silicon, yielding the ridge height diagrammed in Fig. 1hi. Subsequently, we pattern an array of slots—which both serve as phononic mirrors and eventually permit suspension through a wet etch—using electron beam lithography with a positive ZEP520A photoresist. We next develop these features in xylenes and perform another RIE chlorine etch that removes the remainder of the silicon within the slots, exposing the oxide under-cladding. Following the RIE process, we perform a wet etch using 49% hydrofluoric acid, which removes the oxide underneath the waveguide, yielding the suspended hybrid photonic-phononic waveguide diagramed in Fig. 1c. The suspended waveguide device is 2.305 cm long and is continuously suspended by an array of 451 nanoscale tethers.

## 6. IMPORTANT PARAMETERS AND DATA ANALYSIS

All of the parameters in our model are corroborated by independent measurements and simulations (i.e., no fitting parameters). In this section, we list the important parameters of our system and give an account of how each is determined.

| Parameters | Value |
|---|---|
| L | 2.305 cm |
| $v_{g,1}$ | $7.385 \times 10^7$ m/s |
| $v_{g,2}$ | $7.163 \times 10^7$ m/s |
| $\Delta k_{as}$ | $-4.5 \times 10^5$ m$^{-1}$ |
| $\gamma$ | $2\pi \times$ 180 MHz |
| $\Gamma$ | $2\pi \times$ 14.2 MHz |
| G$_b$ | 470 W$^{-1}$m$^{-1}$ |
| $\Omega_b$ | $2\pi \times$ 6.02 GHz |

L, the length of the system, is simply given by the length of the suspended region of the device. The group velocities of the symmetric and antisymmetric optical spatial modes ($v_{g,1}$ and $v_{g,2}$, respectively) are determined through racetrack ring free spectral range (FSR) measurements (see Supplementary Section of Ref. [4]). The wavevector mismatch between the two optical spatial modes ($\Delta k_{as} = k_2(\omega_p + \Omega_b) - k_1(\omega_p)$) was found through finite element simulations. As required by phase-matching, this wavevector mismatch gives the wavevector of the phonon field (i.e., $q = \Delta k_{as}$). $\gamma$, the optical dissipation rate of the antisymmetric optical spatial mode, is given by the linear propagation loss of the antisymmetric optical waveguide mode (i.e., $\gamma = v_g \alpha$).

The acoustic dissipation rate ($\Gamma$) is given by the spectral width of the spontaneous Stokes and anti-Stokes spectra at low pump powers. The Brillouin gain coefficient (G$_b$) was found from nonlinear laser spectroscopy measurements presented in Ref. [3]. These measurements were performed on a device of identical dimensions. $\Omega_b$, the Brillouin frequency, is determined through the heterodyne spectroscopy presented in this work.

In addition to these parameters, we use the nonlinear loss coefficients given in the supplement of Ref. [3] to determine the average on-chip probe and scattered powers.

---



\end{document}